# Carbon-poor stellar cores as supernova progenitors

R. Waldman and Z. Barkat

Racah Institute of Physics, The Hebrew University, Jerusalem 91904, Israel

E-mail: waldman@cc.huji.ac.il

# A B S T R A C T

Exploring stellar models which ignite carbon off-center (in the mass range of about $1.05 - 1.25\ M_\odot$, depending on the carbon mass fraction) we find that they may present an interesting *SN I* progenitor scenario, since whereas in the standard scenario runaway always takes place at the same density of about $2\ X\ 10^9\ gr/cm^3$, in our case, due to the small amount of carbon ignited, we get a whole range of densities from $1\ X\ 10^9$ up to $6\ X\ 10^9\ gr/cm^3$.

These results could contribute in resolving the emerging recognition that at least some diversity among *SNe I* exists, since runaway at various central densities is expected to yield various outcomes in terms of the velocities and composition of the ejecta, which should be modeled and compared to observations.

# 1. Introduction

As soon as they have been identified as a distinct type of supernovae, type Ia supernovae (SN Ia) have proved to have a relatively small dispersion of their luminosity, and hence they were used as distance indicators ("standard candles"), having especially major significance in the effort of determining the cosmological parameters of our universe.

After some years, as the observational database of supernovae increased and became more detailed and accurate, it was recognized that the "standard candle" pictures is inaccurate, since a certain scatter in SN Ia luminosities exists, and several empirical correlations where found, which connect the maximum luminosity with light curve shape, color evolution, spectral appearance, and host galaxy morphology. However, a physical understanding of the origin of this luminosity variation is still lacking. Moreover, some supernovae remain super- or sub-luminous even after these corrections are applied.

On the theoretical side, the almost unanimously accepted explanation of the phenomenon is the explosive burning of degenerate carbon in the core of a carbon – oxygen white dwarf (WD), which becomes unstable as it grows to Chandrasekhar's mass by accretion from a binary companion overflowing its Roche lobe (single degenerate scenario), or a merger of two WD's, following the angular momentum loss from the system by gravitational radiation (double degenerate scenario). However, theoretical models are still far from accurately reproducing crucial features of the observational data, such as the composition of the ejected matter. For a detailed review of the above see e.g. *Leibundgut 2000, Hillebrandt & Niemeyer 2000.*

As for the diversity in the luminosity, several explanations have been suggested, such as variations in the metallicity of the progenitor, in the carbon to oxygen ratio at its center, or in the central density at the time of ignition (*Timmes et al 2003*, *Röpke et al 2006*, *Lesaffre et al 2006*). The variation of latter two parameters is expected to be an outcome of the variation in the initial WD mass and in the accretion history.

In this work we suggest an evolutionary scenario which could lead to a rather unusual, albeit possibly rare type of progenitor, and thus add somewhat to the diversity. We shall look at WD's in the mass range where they undergo an off-center non-degenerate carbon burning episode before the onset of accretion, leaving behind a star mostly, but not completely depleted of carbon, having about 2% carbon in the central region of about 0.5 $M_\odot$. At reaching Chandrasekhar's mass the central density of these progenitors could vary over a comparatively broad range, depending on several factors, as will be discussed later.

Let us first review the standard single degenerate scenario. In this scenario we have a carbon – oxygen white dwarf below Chandrasekhar's mass ($M_{ch}$), which is a remnant of an intermediate mass AGB star, that has lost its hydrogen rich envelope through binary evolution. At a certain stage the white dwarf begins accreting mass from its secondary companion at an "appropriate" rate ($\dot{M}$). The increase in mass raises the density and temperature at the center, so that at a certain stage their values cross the "carbon ignition line"* (hereafter *"IG"*), and carbon is ignited. The immediate outcome is an increase of the entropy at the center, leading to the growth of a convective region. If the size of the convective region reaches a certain value, further increase in entropy leads to expansion of the center (a decrease of the density at the center – hereafter *"DEC"*), in opposing to the accretion tending to lead to contraction. Thus, between *IG* and *DEC* the density at the center increases, and after *DEC* it decreases. The temperature nevertheless continues to increase due to the nuclear reactions, and with it the nuclear reaction rate and the convective flux are also increasing. When the reaction rate reaches a point where the convection can no longer compete with the entropy production rate, nuclear burning continues at almost constant density, and could reach a "dynamic" regime, where the nuclear time (defined as the time needed to exhaust all the fuel, including oxygen, at constant density) is shorter than the dynamic time (which can be defined for example as a pressure scale height divided by the speed of sound). We will refer to this situation as a "runaway" (*RA*), and it is clear that under these circumstances hydrostatic equilibrium can no longer be assumed. It is worth noting, that the runaway doesn't necessarily lead to explosion, since electron capture behind the explosion front could produce a rarefaction wave which might convert the explosion into a collapse. Figure 1 shows the evolution of the density and temperature at the center of a carbon – oxygen star model of mass $M = 1.18\,M_\odot$ and carbon mass fraction of $X_c = 0.05$, displaying also the relevant carbon ignition line and the points *IG*, *DEC* and *RA* mentioned above.

_______________________

* The ignition line is defined as the locus in the $\rho_c$, $T_c$ plane, where the energy production rate from nuclear reactions equals the neutrino losses. We will refer to this point as IG.

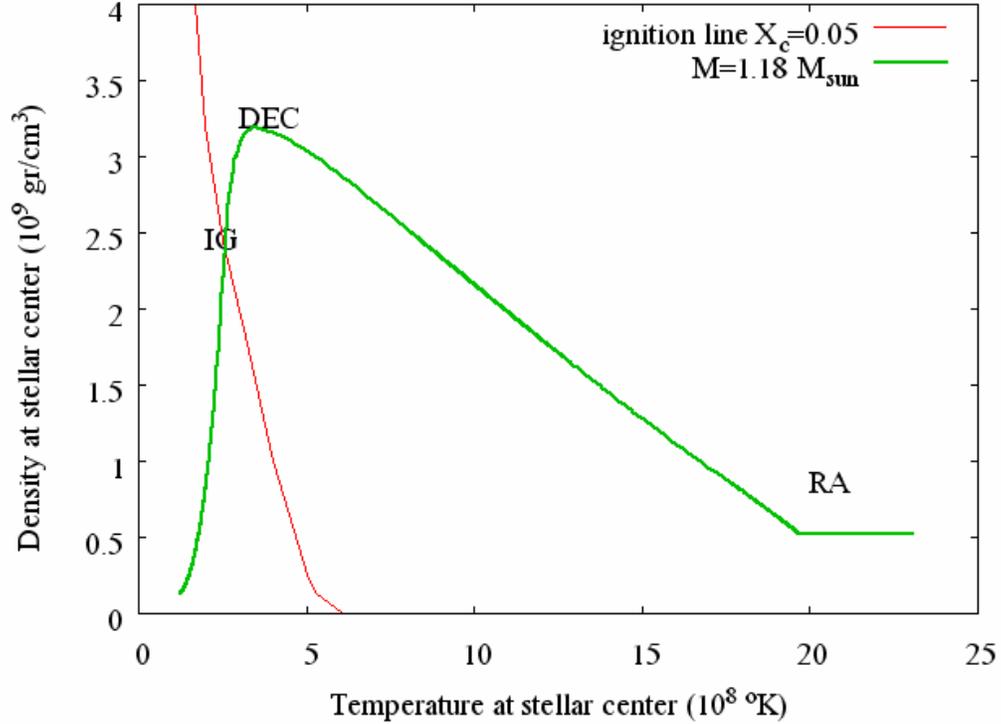

**Figure 1: The evolution of the temperature and density at the center of a typical carbon – oxygen star model, showing the various points along the path from ignition of carbon to runaway.**

Quantitatively speaking, it is clear that the *IG* point is dependent on the carbon mass fraction ($X_c$), and will be located at a higher density for a lower carbon mass fraction. It also depends on the reaction rate, including the screening factor, and on the neutrino loss rate, but the sensitivity to these parameters is weaker. The *DEC* point is also dependent on $X_c$. For a low enough $X_c$ the amount of carbon is insufficient to raise the temperature to the *RA* regime, and even not to the *DEC* point.

If we take into account the Q-value of carbon burning ($\approx 4 \times 10^{17}$ *erg/gr*) and the specific heat in the region under discussion ($(\partial T / \partial e)_\rho \approx 10^{-7}$ *°K/erg*), we can estimate that the temperature will rise by about *4 X 10⁸ °K* for a mass fraction of one percent carbon ($X_c = 0.01$). Since for *RA* the threshold temperature of oxygen ignition should be reached, i.e. about *1.4 X 10⁹ °K* is needed, it is clear that if *IG* is reached at $T_c \approx 4 \times 10^8 \ ^oK$, we need a carbon mass fraction of about *0.025* to reach *RA*. Clearly this is a rough estimate, since part of the energy produced by the burning of carbon is transported away, while on the other hand convection supplies fresh fuel. As we shall see later on, this estimate turns out to be pretty good.

To the abovementioned we should add the role which electron capture processes might play during evolution. A detailed discussion will be given in section 2.3.3; here we shall mention that the relevant processes are:

1. Electron capture on $Mg^{24}$, which is a product of carbon burning and is quite abundant, and $Na^{24}$, which cause a decrease of $Z/A$, and thus a decrease of the effective Chandrasekhar's mass, leading to an accelerated compression and local heating.

2. "*Urca*" processes of two kinds – thermal and convective.

In this work we also checked the influence of these processes.

Regarding the astrophysical scenario, it is usual to deal with white dwarves that are the remnants of planetary nebulae (*PN*) formation. This is a situation when the growth of the carbon – oxygen core of the star causes a luminosity increase leading to envelope instability and ultimately to its ejection.

The typical mass of such white dwarves is about *0.6 $M_\odot$,* although more massive ones do exist. According to *Liebert et al. 2005*, some 6% of the white dwarves have masses above *1 $M_\odot$*. Regarding their composition, the carbon mass fraction generally lies within the range *0.25 ≤ $X_c$ ≤ 0.55*, due to some uncertainties.

As mentioned before, the key point of the standard scenario is mass accretion from a binary companion. Clearly, the accretion rate depends on the structure and evolutionary history of the binary couple. Theoretical surveys have been made in the literature, where the accretion rate, as well as the initial mass of the accreting white dwarf and the composition of the accreted matter served as free parameters (e.g. *Nomoto & Sugimoto 1977*).

In this work, we referred to the WD mass range where the original carbon core is big enough to ignite carbon before it grows towards $M_{ch}$, but not too big so that ignition takes place off-center. Alas, investigating the required mass range is especially difficult, due to the need for extremely fine zoning in order to follow the burning shell and the convective region created above it. In section 3.2 we discuss this subject in detail.

In the following, we shall begin by first describing the computational methodology of our work, including the numerical algorithms and input physics in chapter 2. Our results are given in chapter 3. Finally, in chapter 4 we will discuss our results and present our conclusions, including observational predictions and suggestions for further research.

# 2. Method

## 2.1. The initial models

For the sake of simplicity, rather than following the complicated evolution of the primary star in order to get the remaining carbon – oxygen white dwarf, we began our calculation with a carbon – oxygen star of appropriate mass and composition, with low enough central density at hydrostatic equilibrium, letting it contract along the well known knee-shaped path on the central density versus central temperature diagram. This approach, which has been used by many authors since, was justified by *Barkat 1971*, stating that at each point along the evolutionary track of a carbon – oxygen core, growing as a result of a burning shell, the central conditions (density and temperature) are very close to the evolutionary track of a carbon – oxygen star of corresponding mass.

## 2.2. Stellar evolution code

The evolution was followed using a quasi-static *Lagrangian 1D* evolution code named *ASTRA*, which is an extensively improved version of the *ASTRA* evolution code first described by *Rakavy, Shaviv & Zinamon 1966*, and used with some modifications many times. The quasi-static assumption, meaning that the star is dynamically stable, and the time scales for the three main processes that govern the evolution of the star – namely the hydrodynamic motion, the convective mixing, and the exchange of energy due to thermonuclear reactions and radiative transport obey the relation: $t_{hydro} << t_{convec} << t_{thermo}$, is valid for most stages of stellar evolution, with the possible exception of very violent nuclear burning phases, where special care has to be taken. Since these stages are indeed in the scope of our interest, a special treatment of these stages was devised, and will be described in section 2.4.

Convective regions are treated as isentropic and fully mixed. This assumption greatly simplifies the program, as it eliminates the need to calculate the convective energy flux, and is valid (i.e. is in good agreement with the mixing length model of convection) wherever the pressure scale height is large compared to the size of the convective region of interest. In the stellar models of our interest, this kind of condition occurs throughout the stellar core.

## 2.3. Input physics

### 2.3.1. Nuclear reaction rates

Nuclear reactions were treated via two different sets of reaction rates, which were then compared to each other:

1. Using an α-network of 13 elements from $He^4$ up to $Ni^{56}$, with reaction rates based on the "NON-SMOKER" tables by *Rauscher & Thielemann 2001*.

2. Using the reaction rates according to *Caughlan & Fowler 1988* for multiple α-nuclei.

The screening factor was treated according to *Itoh et al. 1979* and *Itoh et al. 1980*, but a comparison was made with those of *Dewitt et al 1973*.

### 2.3.2. Neutrino losses

Neutrino losses were taken according to *Itoh & Kohyama 1983*.

### 2.3.3.    Electron capture

Three different processes of electron capture are considered:

#### 2.3.3.1.    *Thermal Urca (TU)*

Thermal *Urca* is a situation where at some *Lagrangian* mass point in the star ($m_{urc}$) the Fermi energy fulfills $E_f = E_{th}$, so that throughout a mass shell (henceforth "*Urca* shell" - *US*), lying in the range of $E_f = E_{th} \pm 1\ kT$, processes of emission and capture of electrons by some trace nuclei take place. Both processes are accompanied by production of neutrinos, which leave the star and create an effective local heat sink. Since the star continues to shrink and $E_f$ increases, it is clear that $m_{urc}$ increases and the *US* moves outward. The magnitude of the process depends on the mass fraction of suitable nuclei (*Ergma & Paczynski 1974*). In our context, the dominant nuclei are $Na^{23}$, with a threshold of $E_{th} = 4.4\ MeV$ and consequently $\rho_{th} \approx 1.7\ X\ 10^9\ gr/cm^3$, and afterwards $Ne^{21}$, with a threshold of $E_{th} = 5.7\ MeV$, and thus $\rho_{th} \approx 3.5\ X\ 10^9\ gr/cm^3$.

We modeled the effect of the *TU* by means of the prescription by *Tsuruta & Cameron 1970*, while checking the sensitivity of the result to the reaction rate by varying the mass fraction of the relevant nuclei.

#### 2.3.3.2.    *Convective Urca (CU)*

In this case convection transfers nuclei through the *US*. These nuclei pass through a region where the difference $E_f - E_{th}$ is of the order of magnitude below *1 MeV*, and certainly $\Delta E >> kT$. Since in this case the average electron capture occurs below the Fermi level, a hole is created, which is filled by an electron from above the Fermi level, and the energy surplus is emitted as a $\gamma$ photon causing local heating.

The effect of *CU* on the relevant evolution has been discussed for a long time in the literature with contradictory conclusions (e.g. *Paczynski 1972, Bruenn 1973, Couch & Arnett 1975, Lazareff 1975, Regev & Shaviv 1975, Barkat & Wheeler 1990, Mochkovitch 1996, Lesaffre et al. 2006*). Lately it turned out that the main effect of *CU* is to hinder the extension of the convective region above $m_{urc}$ (*Stein et al. 1999, Bisnovati-Kogan 2001*). Hence, in this work we tried to examine the practical significance of *CU*, using a simplistic approach, by artificially forbidding convection above $m_{urc}$.

### 2.3.3.3. *Electron capture on carbon burning products*

As *Miyaji et al. 1980* have already shown, processes of electron capture by certain burning product nuclei, when the Fermi level ($E_f$) exceeds a relevant threshold ($E_{th}$), can have a major importance during the stage preceding *RA*, when the density, and consequently the Fermi energy, are rising. In our context, the dominant nuclei from among the carbon burning products are\* (*Miyaji et al. 1980* again) $Mg^{24}$, with a threshold of $E_{th} = 5.52\ MeV$ and consequently $\rho_{th} \approx 3.35\ X\ 10^9\ gr/cm^3$, and afterwards $Na^{24}$, which is the electron capture product of $Mg^{24}$, with a threshold of $E_{th} = 6.59\ MeV$, and thus $\rho_{th} \approx 5.25\ X\ 10^9\ gr/cm^3$.

Note that lately (*Gutierrez et al. 2005*) there have been claims that $Na^{23}$ is much more abundant than previously thought, and is more abundant than even $Mg^{24}$. Since the threshold for *EC* on $Na^{23}$ is lower than on $Mg^{24}$, it should have been considered, however *Gutierrez et al. 2005* have shown that the effect on the evolution is small.

---

\* Albeit the mass fraction of $Ne^{20}$ in the burning products of carbon is high, we didn't need to include *EC* on it, since the threshold in its case is high, and corresponds to $\rho > 6\ X\ 10^9\ gr/cm^3$, which is above the limit of our interest.

In order to calculate the effect of *EC*, it is necessary to know the mass fraction of the relevant nuclei, which is a result of carbon burning and the capture rate, as well as the neutrino loss rate and the resulting heating rate.

As will be evident from our results, this process has an important effect in our cases; therefore it was included as a standard in all our models, except when explicitly mentioned otherwise. Due to the existing uncertainties (cf. *Gutierrez et al. 2005*), in line with our basic approach, we checked the sensitivity of the results both to the mass fraction and to the capture rate by introducing "fudge factors".

We took the capture rates from *Miyaji et al. 1980*, but the range of variations we checked covers also the differences versus the rates given by *Oda et al. 1994*, which differ from the former by as much as an order of magnitude.

*Mochkovitch 1984* pointed out that in the presence of *EC*, due to its effect on the gradient of the electron mole number $Y_e$, the *Ledoux* criterion for convection has to be used, and hence the extension of the convective region is slower. In our case it turned out, that the effect of using the *Ledoux* criterion is quite small.

Another important caveat (*Stein 2005*): since it is clear that *EC* must start as *TU* (which is locally cooling), and can become exothermic only when occurring below the threshold by more than *kT*, the question whether a convective zone broad enough to ensure heating can be formed arises. In our case, where *EC* starts when significant carbon burning is already present, this problem may not be severe. However, more careful analysis is needed, to find out whether the local cooling might force the base of convection to move outward.

### 2.3.4. Opacities

Radiative opacities were calculated according to the *OPAL* opacity tables (see *Iglesias & Rogers 1996*), using the tables and interpolation subroutine provided at the web site *www-phys.llnl.gov/Research/OPAL*. The tables were extended to lower temperatures according to *Alexander & Ferguson 1994*.

Electron conduction was calculated according to *Iben 1975*.

### 2.3.5. Equation of state

The equation of state takes into account ionizations to all available levels of the different atom species in the composition. The distribution function of the various ionization levels is computed using a method similar to the one described in *Kovetz & Shaviv 1994*. The resulting electron density is then used together with the temperature in order to extract the pressure, energy, chemical potential and their derivatives respective to the electron density and temperature from a table computed in advance by solving the *Fermi-Dirac* integrals. The pressure, energy and entropy of the ions are then added as an ideal gas together with those of the radiation.

## 2.4. Treatment of deviations from the quasi-static assumption

As previously mentioned, our evolution code assumes, that the time scales for the three main processes that govern the evolution of the star – namely the hydrodynamic motion, the convective mixing, and the exchange of energy due to thermonuclear reactions and radiative transport, obey the relation: $t_{hydro} \ll t_{convec} \ll t_{thermo}$.

However, as we shall see later, as the star approaches explosion, this relation between the timescales is no longer valid. As the reaction timescale $t_{thermo}$ becomes shorter, it first of all becomes comparable to the convective timescale $t_{convec}$, so in order to achieve a reasonable modeling, we have to appropriately restrict convection when this occurs. Hence, it is important to estimate these two timescales, and give an adequate treatment when the quasi-static assumption fails. In practice, as a very crude model, we restrict the outer boundary of the convective zone (the inner boundary in our case is always at the center) to be not above the innermost *Lagrangian* zone, which fulfills $t_{convec}(r) > \alpha \, t_{thermo}$, where $r$ is the radius of the zone, and $\alpha$ is a "fudge factor" we use to check the sensitivity of the results. Note that $1/\alpha$ is in fact the number of convective turnover times during an interval of one thermonuclear timescale. The method for calculating the timescales $t_{thermo}$ and $t_{convec}$ is given below.

### 2.4.1. The convective timescale

The widely used mixing length theory gives an explicit relation between the convective luminosity and the convective velocity:

*(2-1)*
$$v_c = \left[ \frac{\lambda P \left( \frac{\partial \ln \rho}{\partial \ln T} \right)_P}{c_P T \rho^2} L_c \right]^{1/3}$$

Here $\lambda$ is the mixing length, $c_p = (\partial e/\partial T)_p$ is the specific heat at constant pressure, and $L_c$ is the convective flux through a unit of area.

We don't use the mixing length theory, but rather assume an isentropic convective region. Nevertheless, we can estimate the convective luminosity as follows:

Let $L_c$ be the convective luminosity and $L_r$ the radiative luminosity. At any point we have:

(2-2)
$$T\frac{\partial S}{\partial t} = q - \frac{\partial L_r}{\partial m} - \frac{\partial L_c}{\partial m}$$

However, we assume that $\partial s/\partial t$ is uniform throughout the convective region, thus:

(2-3)
$$\int T\frac{\partial S}{\partial t}dm = \frac{\partial S}{\partial t}\int Tdm = \int qdm - \Delta L_r - \Delta L_c$$

But $\Delta L_c = 0$ since the convective flux vanishes at the boundaries of the convective region, so we have:

(2-4)
$$\frac{\partial S}{\partial t} = \frac{\int qdm - \Delta L_r}{\int Tdm}$$

Together with (2-2) we have:

(2-5)
$$\frac{\partial L_c}{\partial m} = q - \frac{\partial L_r}{\partial m} - T\frac{\int qdm - \Delta L_r}{\int Tdm}$$

From the above equation we can derive the convective luminosity $L_c(m)$ for each point in the convective region, and then use (2-1) to get the convective velocity $v_c(m)$ at each point in the region.

The time needed for the convective flux to cover a distance $\Delta r$ in the vicinity of a *Lagrangian* point $m$ is of course:

(2-6)
$$\Delta t = \frac{\Delta r(m)}{v_c(m)}$$

Consequently the time to reach a distance $r$ from the inner boundary of the convective region will be:

$$\text{(2-7)} \qquad \tau(r) = \int dt = \int \frac{dr}{v_c}$$

The "convective time scale" will be defined as the time $\tau(r)$ to the outer boundary of the convective region, i.e. the time needed to cross the entire length of the region.

It is clear that this is only a rough estimate; nevertheless a comparison with models of convective envelopes calculated for the same conditions with the mixing length theory using the method described by *Tuchman et al. 1978* gave an agreement better than a factor of 3.

### 2.4.2. The nuclear timescale

As the nuclear timescale we define the time needed to burn all the carbon and afterwards the oxygen at a given point (with initial temperature $T$, density $\rho$ and carbon mass fraction $X_c$), under the assumption of constant density. Clearly this is a lower limit to the time, since in reality there are also energy losses.

To get a quantitative estimate we used the analytical approximation of *Woosley & Weaver 1986* for the reaction rates of carbon and oxygen burning in the relevant range (i.e. *$2 < T_9 < 6$, $\rho_9 < 4$*):

$$\text{(2-8)} \qquad \begin{aligned} \text{carbon}: \ & q_c(\rho, T) = 8.25 \times 10^{15} \, X_c^2 \, \rho_9^{2.79} T_9^{22} \ \text{(erg/sec)} \\ \text{oxygen}: \ & q_o(\rho, T) = 8.25 \times 10^{15} \, X_o^2 \, \rho_9^{2.79} T_9^{36} \ \text{(erg/sec)} \end{aligned}$$

They also approximated the heat capacity using:

(2-9)
$$\frac{\partial T}{\partial e} = 1.57 \times 10^{16} \, \rho_9^{-0.26} T_9^{0.76}$$

Combining the last two equations we get (for carbon):

(2-10)
$$\frac{\partial T}{\partial t} = q \frac{\partial T}{\partial e} = 1.3 \times 10^{32} \, X_c^2 \rho_9^{2.53} T_9^{22.76}$$

A similar equation can be obtained for the oxygen. Accordingly, the time for raising the temperature from $T_1$ to $T_2$ is given by:

(2-11)
$$t_{1 \to 2} = \frac{1}{1.3 \times 10^{32} \, X_c^2 \rho_9^{2.53}} \int_{T_1}^{T_2} \frac{dT}{T_9^{22.76}}$$

Given an initial temperature $T_1$, we can estimate the final temperature $T_2$ of carbon burnout, since:

(2-12)
$$\int_{T_1}^{T_2} \left( \frac{\partial e}{\partial T} \right)_\rho dT = X_c Q_v \qquad (Q_v \text{ is the Q-value})$$

After we get the time $t_{1 \to 2}$ for carbon burnout and the final temperature $T_2$, we use $T_2$ as the initial temperature of oxygen burning, and through the same method we can get the time $t_{2 \to 3}$ of oxygen burnout.

Note that we do not take into account the fact, that $X_c$ is also a function of time, but it turns out to have a small effect. In fact, in our calculations we checked the validity of these estimates by artificially preventing the density from changing, and actually measuring these times. We found excellent agreement.

Estimating the timescale of oxygen burning is important, since in case the carbon mass fraction is very low, it might burn out without igniting the oxygen.

# 3. Results

## 3.1. Overview

The evolution of carbon – oxygen stars is determined by their mass ($M_c$) and composition. We assume the composition of the star is homogeneous – as a result of the helium burning producing it being convective. Therefore it is sufficient to specify the mass fractions of the relevant elements: $C^{12}$ and $O^{16}$, whereas the mass fractions of *Ne* and *Mg* are negligible. These mass fractions are determined at the end of helium burning, for which there is a well known uncertainty, due to uncertainties in the cross-section for the reaction $C(\alpha,\gamma)O$, which takes place during helium burning. It is common to assume the mass fraction of the carbon ranges between *$0.25 \leq X_c \leq 0.55$*, and the mass fraction of the oxygen is its complementary to unity (*Umeda et al. 1998*).

We can identify four fundamentally different ranges. Above Chandrasekhar's mass ($M_3 = M_{ch} \approx 1.4 M_{\odot}$) carbon is ignited at the center, and the stellar center evolves toward increasing temperatures and densities, igniting heavier fuels.

Below *$M_3$*, but above a certain limit *$M_2$*, carbon is ignited at the center, but no heavier fuels are ignited, and finally evolution proceeds towards a white dwarf. The value of *$M_2$* depends on the carbon mass fraction; for our example of *$X_c = 0.54$* it lies around *$M_2 \approx 1.17 M_{\odot}$*

Below *$M_2$*, but above a lower limit *$M_1$*, carbon is ignited off-center, at a point depending both on the mass of the star and on the carbon mass fraction (see Figure 3 and Figure 4). Subsequently, carbon burning propagates both inwards toward the center and outwards, and after almost all the carbon in the core is exhausted, evolution returns to its original path towards a white dwarf.

As we already suggested in chapter 1, and will show in detail later, these stars are in the focus of our interest, since the off-center burning might leave behind enough carbon, which will "survive" the subsequent evolutionary phases, and finally ignite explosively after the mass reaches $M_{ch}$ through accretion. Figure 2 shows a typical example of an off-center carbon igniting star with mass of $M = 1.17\ M_{\odot}$.

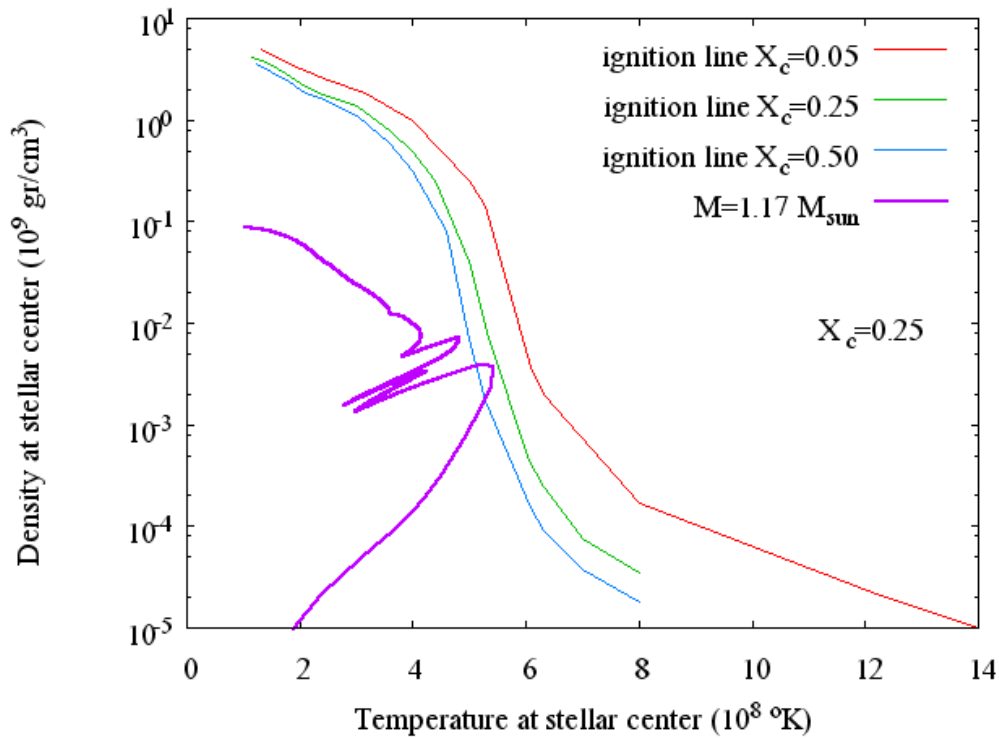

**Figure 2: The evolution of the density vs. the temperature at the center of a 1.17 M$_{\odot}$ carbon star model, which ignites carbon off center (M$_1$ < M < M$_2$). The carbon mass fraction is X$_c$ = 0.25. The ignition lines are also plotted for various carbon mass fractions.**

We can see that at a the point of ignition the density and temperature at the center decrease, as a result of the expansion induced by the carbon burning shell above. As we will explain in detail in section 3.2.3, this shell extinguishes, subsequently causing the center to contract and heat again, but soon another off-center burning shell is ignited, causing the center to expand and cool a second time. Finally the burning reaches the center, causing it to rise to carbon burning temperature with almost no change in density. After the burning ceases, the center returns to its original path, as if no carbon burning had taken place.

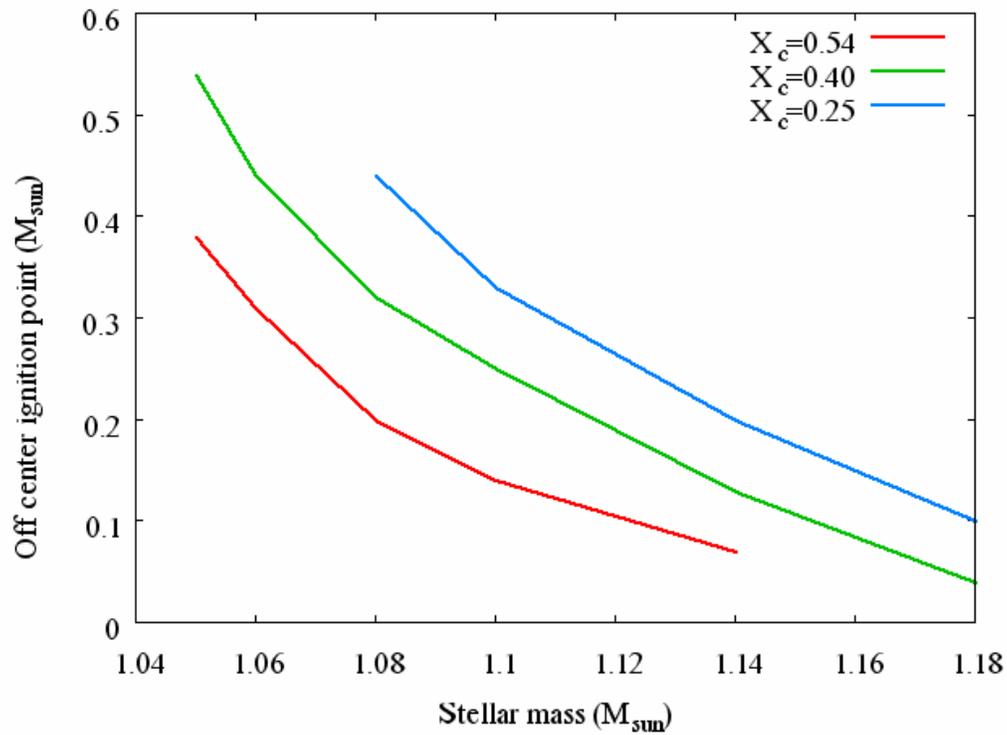

**Figure 3: The off-center ignition point vs. the stellar mass for various carbon mass fractions.**

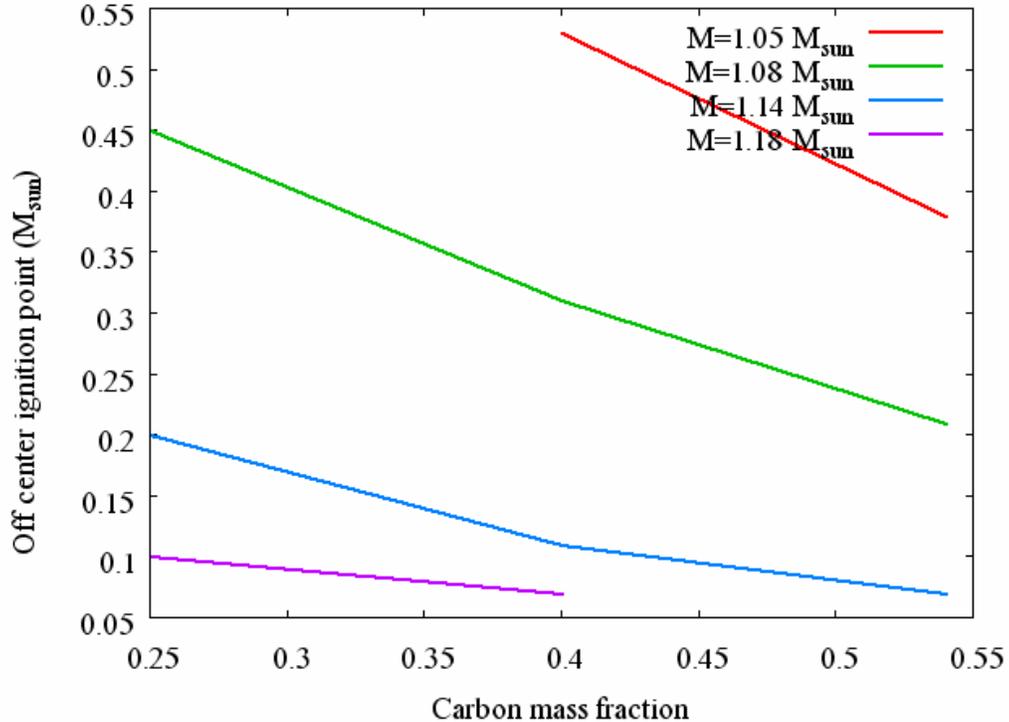

**Figure 4: The off-center ignition point vs. carbon mass fraction for various stellar masses.**

Below the limit $M_1$ the star evolves towards a white dwarf without the carbon being ignited. The value of $M_1$ depends of course on the carbon mass fraction. For a carbon mass fraction of $X_c = 0.54$, we have $0.95\ M_\odot < M_1 < 1.05\ M_\odot$, while for a lower carbon mass fraction of $X_c = 0.25$, the limit is slightly higher standing at $1.05\ M_\odot < M_1 < 1.08\ M_\odot$.

### 3.2. Off-center carbon burning

We will limit our interest to the range $M_1 \leq M_c \leq M_3$. In order to point out the major points of interest in this range, we will first describe a specific example.

We will take as an example a model with mass $M = 1.17\ M_\odot$ and a homogeneous mass fraction profile, i.e. $X_c(m) = X_c(0) = 0.25$. We will investigate the evolution by looking at various physical quantities. Figure 5 presents a Kippenhahn diagram, which shows the history of convective regions in the star. As can be seen, the star goes through various evolutionary stages, each one presenting a burning shell topped by a convective region.

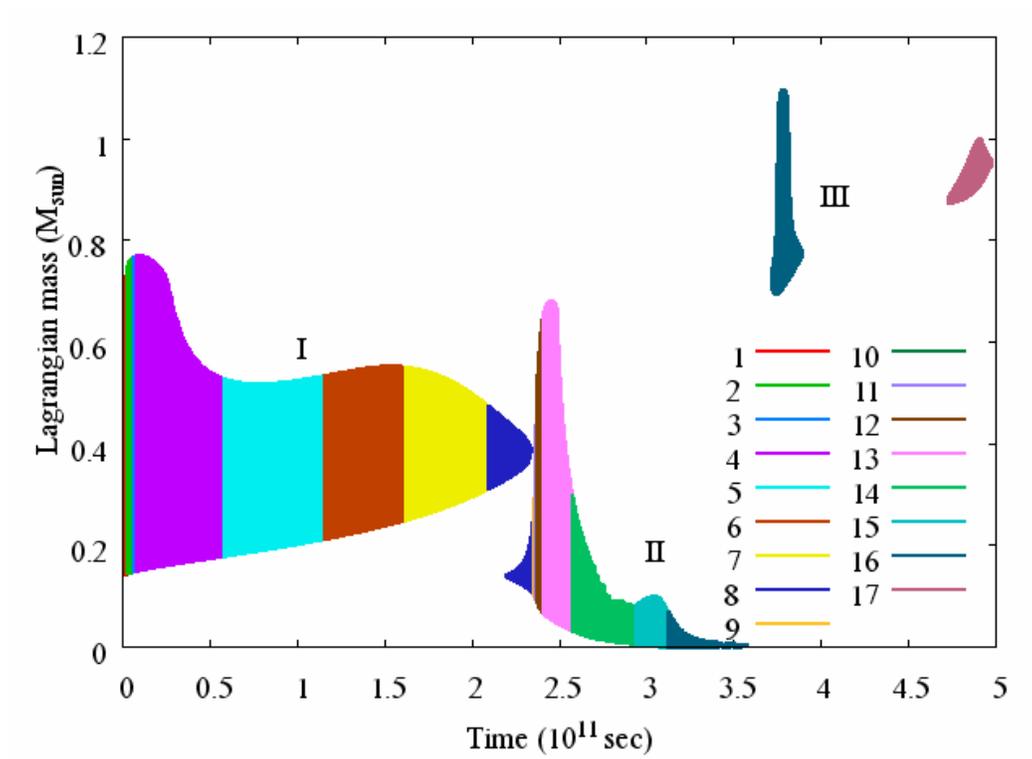

**Figure 5: The history of the convective regions during off-center carbon burning in a 1.17 M$_\odot$ carbon oxygen model with initial carbon mass fraction of X$_c$ = 0.25. The numbers I, II, III refer to the various stages of burning referred to in the text below. The colors correspond to the colors of the profiles in the following figures.**

We will describe the three burning stages shown above, by following the evolution of the carbon mass fraction profile.

### 3.2.1. Stage I

The evolution of the carbon mass fraction is shown* in Figure 6, and we can see that in this case carbon ignites off-center, at $m \approx 0.145\ M_\odot$. The nuclear reaction rate ($q_n$) grows rapidly, and when it exceeds the energy loss rate by neutrino emission ($q_\nu$), a gradually growing convective burning zone develops, while the carbon mass fraction $X_c$ gradually decreases. At a certain stage (line 4 in Figure 6) the convective region begins to shrink, by retreating of both its inner and outer boundaries. This evidently leaves behind a gradient in the carbon mass fraction $X_c$. It is notable, that immediately below the inner boundary of the convective region lies a narrow radiative burning shell, which locally exhausts the carbon at a relatively higher rate. This shell slightly penetrates inward due to conduction. The decline of $X_c$, together with the expansion caused by the rise in entropy of the convective region, finally extinguishes $q_n$. It is important to realize, that due to $q_\nu$, the nuclear burning $q_n$ extinguishes, albeit $X_c$ has not completely vanished. This is a major point that will have important repercussions in what follows.

---

\* Note that *Gutierrez et al. 2005* has a very similar figure.

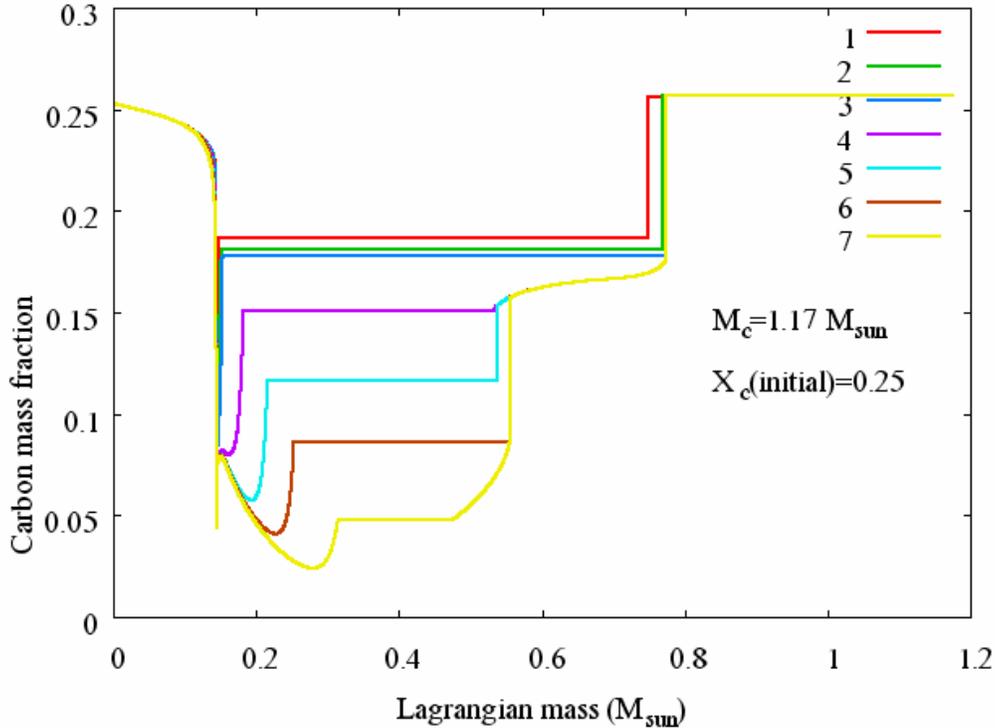

**Figure 6: The evolution of the carbon mass fraction profile in a 1.17 $M_\odot$ mass carbon star model during the stage I of off-center carbon burning. Each line represents the mass fraction profile at a different point in time, in the order of numbering in the legend, which corresponds to the numbering in Figure 5.**

### 3.2.2. Stage II

After the nuclear burning extinguishes, the star continues to contract, leading momentarily to ignition of carbon in the region of the $X_c$ gradient remaining at the base of the former convective burning zone. Again $q_n$ rapidly rises above $q_\nu$, and a burning zone penetrates inward forming a convective region above it. The evolution of the carbon mass fraction during this stage is shown in Figure 7.

The behavior of $X_c$ in the convective region is complex. Although nuclear burning obviously decreases $X_c$, following the growth of the convective region, variations of $X_c$ due to incorporation of zones richer or poorer in carbon have to be taken into account. And indeed at certain stages $X_c$ in the convective region increases, while at other stages it decreases. This behavior varies from star to star, since it is dependent on the details of the preceding evolutionary stage. It repeats itself several times during subsequent evolutionary stages, and cannot be described in general terms, or as a phenomenon monotonically dependent on the stellar mass or the initial carbon mass fraction $X_c(t=0)$.

Several authors have described the case of an inward advancing carbon burning shell topped by a convective region. *Kawai et al. 1987* followed a carbon burning shell that has been ignited at *m = 1.07 $M_\odot$*, while mass was accreted on the white dwarf at a rate of *2.7 X 10$^{-6}$ $M_\odot$/yr*, while *Saio & Nomoto 1998* found a similar ignition at *m = 1.04 $M_\odot$* for an accretion rate of *1 X 10$^{-5}$ $M_\odot$/yr*. As already mentioned, the conclusion of these two papers is, that the shell penetrates inward only up to a certain point, where it diminishes as a result of the expansion induced by the rise of entropy in the convective region above the shell. After extinction of the shell, the star contracts, and a new shell is ignited at the point where the former one had extinguished, this possibly reoccurring several times. We didn't explicitly interest ourselves in shells ignited far from the center, since in our case the waning of the shell which ignites far from the center is a result of the lack of fuel depleted by previous burning episodes. However, also in the case described in section 3.2.1 as "stage I" the expansion contributes to extinguishing the shell and especially its radiative leading part. We will mention that in "stage II" the expansion is minimal since the extent of the convective region is small.

*Timmes et al. 1994* took the effort to estimate, based on reasonable assumptions, the velocity of the shell as a function of the density, temperature and carbon mass fraction, giving a tabulation of their results. A comparison with our results reveals a good agreement. For a case where $\rho = 6.4 \; X \; 10^5 \; gr/cm^3$, $T = 7.3 \; X \; 10^8 \; ^oK$, $X_c = 0.32$, the velocity of the shell is about $1.3 \; X \; 10^{-3} \; cm/sec$. A look at the relevant table in *Timmes et al. 1994* shows a reasonable agreement, since for $X_c = 0.30$ they have a velocity between $8.47 \; X \; 10^{-4} \; cm/sec$ (for $T = 7 \; X \; 10^8 \; ^oK$) and $6.10 \; X \; 10^{-3} \; cm/sec$ (for $T = 8 \; X \; 10^8 \; ^oK$). In the work of *Gil-Pons & Garcia-Berro 2001*, dealing with the formation of oxygen – neon dwarves during mass exchange in binaries, as well as in the series of papers by *Garcia-Berro & Iben 1994*, *Ritossa, Garcia-Berro & Iben 1996*, *Garcia-Berro, Ritossa & Iben 1997*, *Iben, Ritossa & Garcia-Berro 1997*, *Ritossa, Garcia-Berro & Iben 1999*, exploring the evolution of intermediate mass stars between 9 – 11 M$_\odot$, at a certain stage carbon is ignited off-center in a manner very much similar to that expected in our case, and even the resulting carbon profile is very much like ours (see e.g. figure 12 in *Gil-Pons & Garcia-Berro 2001*).

After the nuclear burning flame reaches the center of the star (line 15 in Figure 7), the reaction rate starts to weaken with the decrease of $X_c$ and the reduction of the convective region.

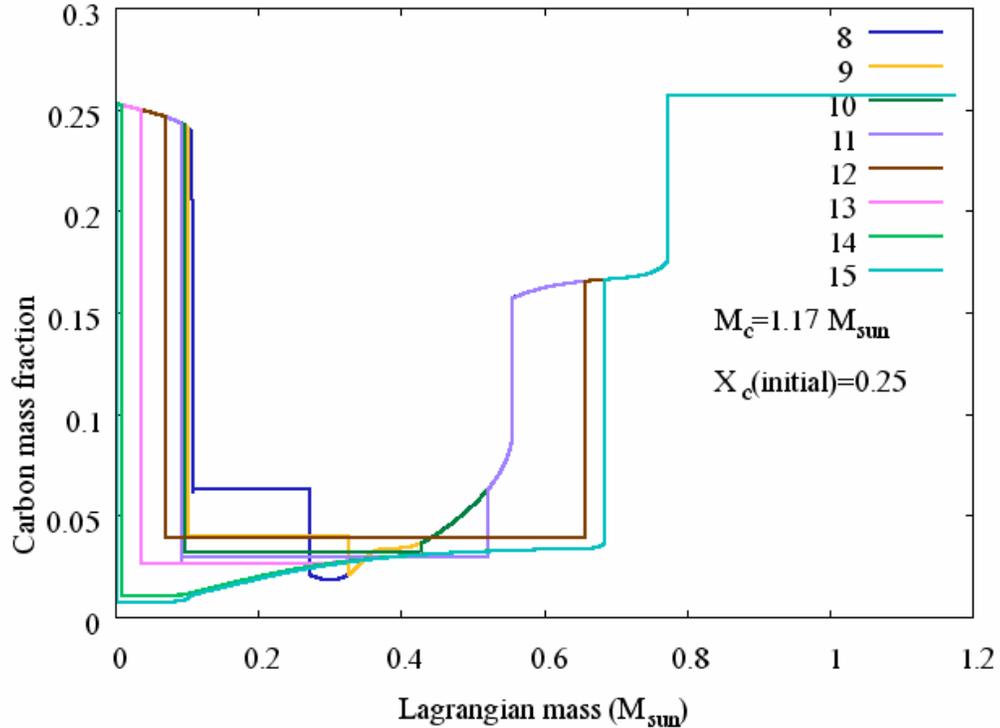

**Figure 7: The evolution of the carbon mass fraction profile in a 1.17 M$_\odot$ mass carbon star model during stage II of off-center carbon burning. Each line represents the mass fraction profile at a different point in time, in the order of numbering in the legend, which corresponds to the numbering in Figure 5.**

Figure 8 displays the advance of the flame inwards, and it is seen that it behaves like a regular almost self similar front. Note however, that close to the center the flame width narrows, and thus special care must be taken, otherwise the flame might incorrectly die out leaving an inner zone of unburned carbon.

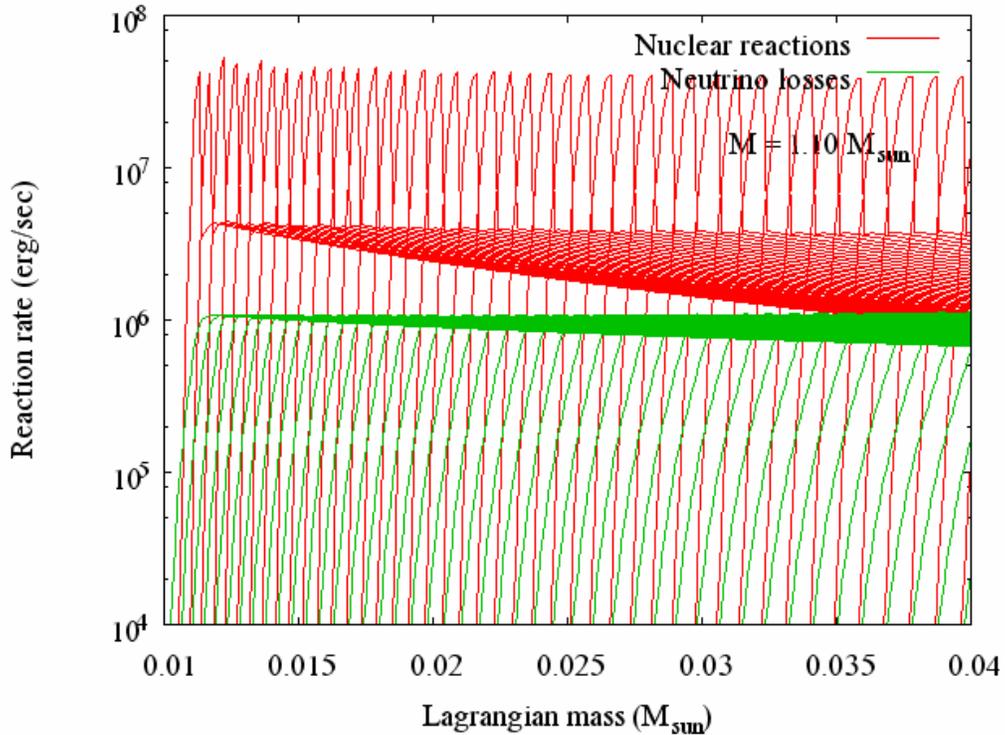

**Figure 8: The inward advancing carbon burning front during "stage II" in a 1.10 M$_\odot$ mass carbon – oxygen model. The nuclear reaction rate (red) and neutrino loss rate (green) are shown at various times.**

### 3.2.3.  Stage III

In the following stage, as a result of the contraction of the star, the carbon reignites, usually in a relatively carbon-rich zone, above the extent of the convective regions of the previous stages.

Also in this case a convective region develops, which grows up to a certain extent, and then the reactions extinguish again. At the same time a radiative burning front develops, which advances inward to the carbon-poor region along the composition gradient which has remained there due to the previous burning. We find (see Figure 9) that this flame gradually decays, and is only able to penetrate slightly.

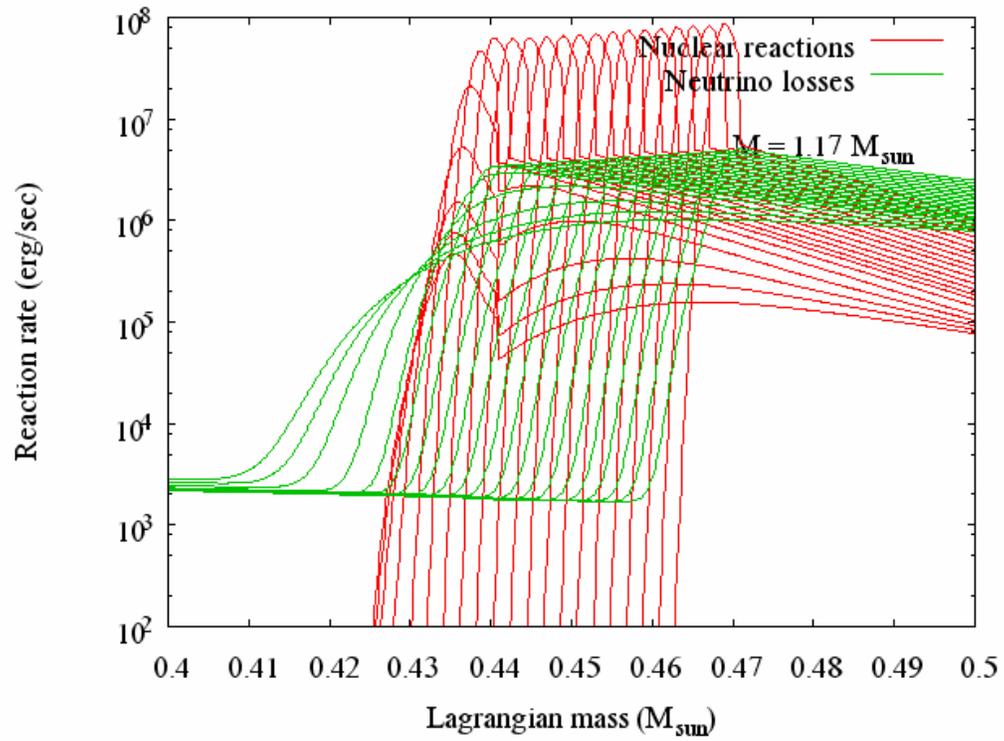

**Figure 9: The decay of the inward advancing carbon burning front during "stage III" in a 1.17 M$_\odot$ mass carbon – oxygen model. The nuclear reaction rate (red) and neutrino loss rate (green) are shown at various times.**

The evolution of the carbon mass fraction during this stage is shown in Figure 10.

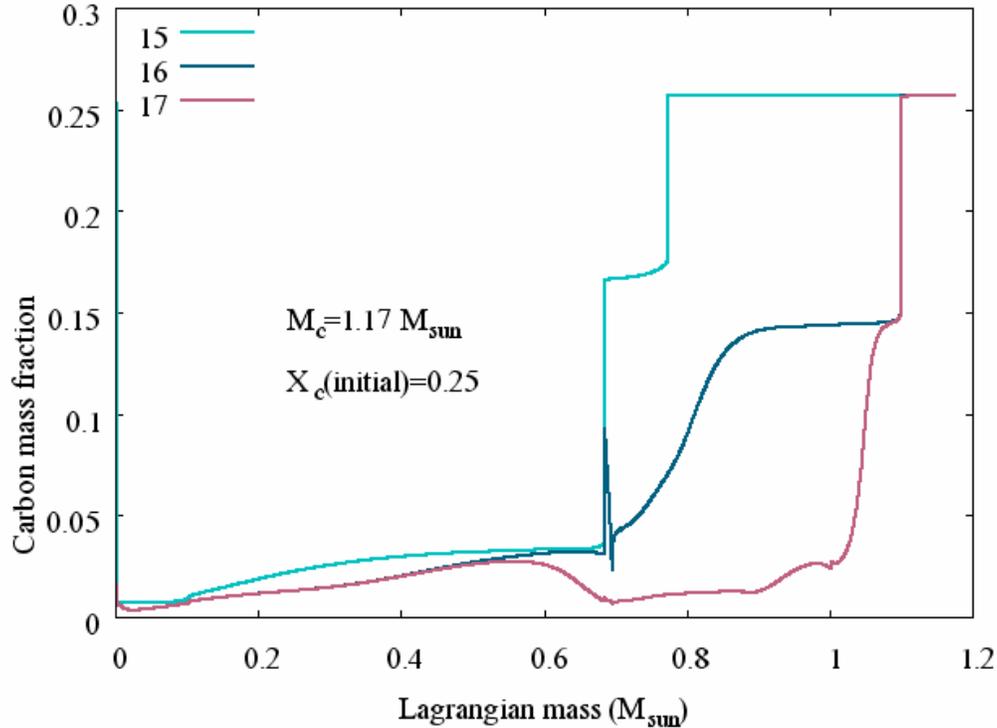

**Figure 10: The evolution of the carbon mass fraction profile in a 1.17 M$_\odot$ mass carbon star model during stage III of off-center carbon burning. Each line represents the mass fraction profile at a different point in time, in the order of numbering in the legend, which corresponds to the numbering in Figure 5.**

In some cases there are more stages where carbon is ignited in carbon-rich outer areas (Figure 5). Eventually, further contraction doesn't result in carbon ignition, and the star evolves toward a white dwarf, with a "frozen" carbon profile. Figure 11 displays this final profile for the main elements present in the star – carbon, oxygen, neon and magnesium.

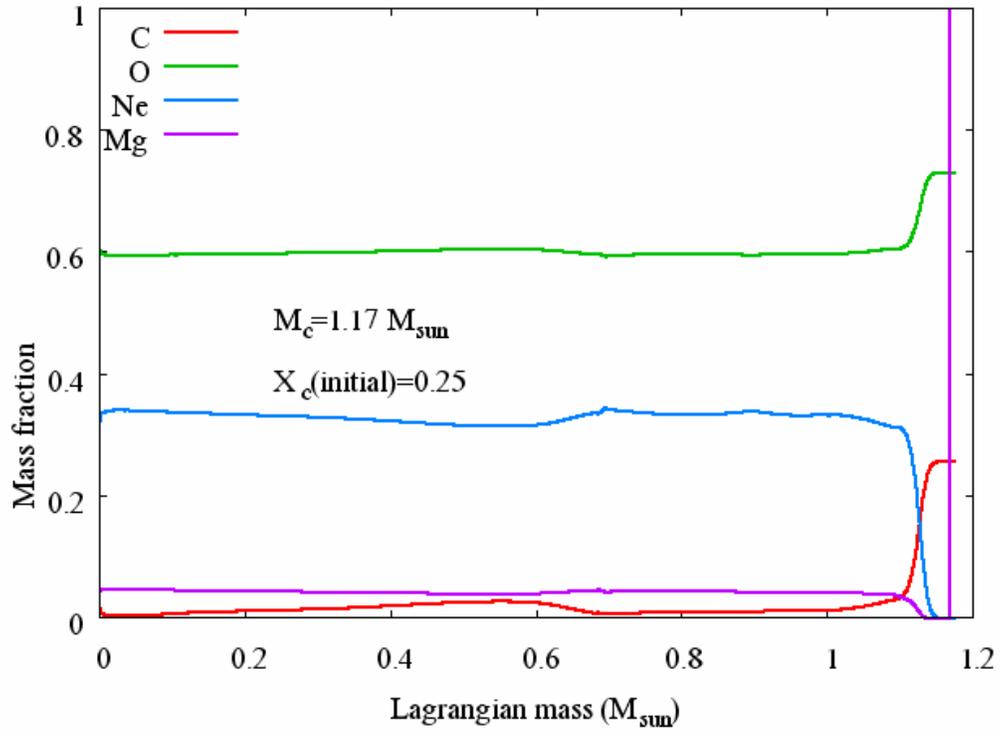

**Figure 11: The final element mass fraction profile in a 1.17 M$_{\odot}$ mass carbon star model after off-center carbon burning.**

Figure 12 displays* the carbon mass fraction profile at the end of carbon burning for various masses and initial carbon mass fraction of *X$_c$(t=0) = 0.54*. Figure 13 displays the carbon mass fraction profile at the end of carbon burning for various initial carbon mass fractions and total mass of *M$_c$ = 1.22 M$_{\odot}$*. We can see that the typical profile has a "bump" around the center, above it a region almost devoid of carbon, and in most cases a small carbon-rich zone near the outer boundary.

---

* Note the change of scale.

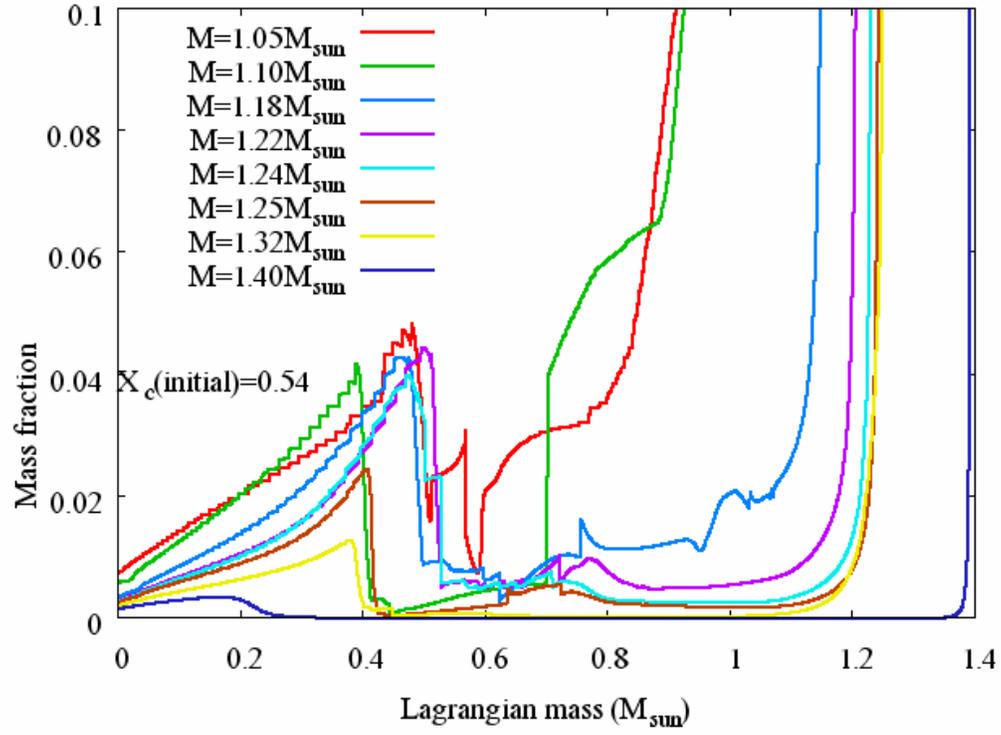

**Figure 12: The final carbon mass fraction profile in carbon star models of various masses and initial carbon mass fraction of $X_c$=0.54 after off-center carbon burning.**

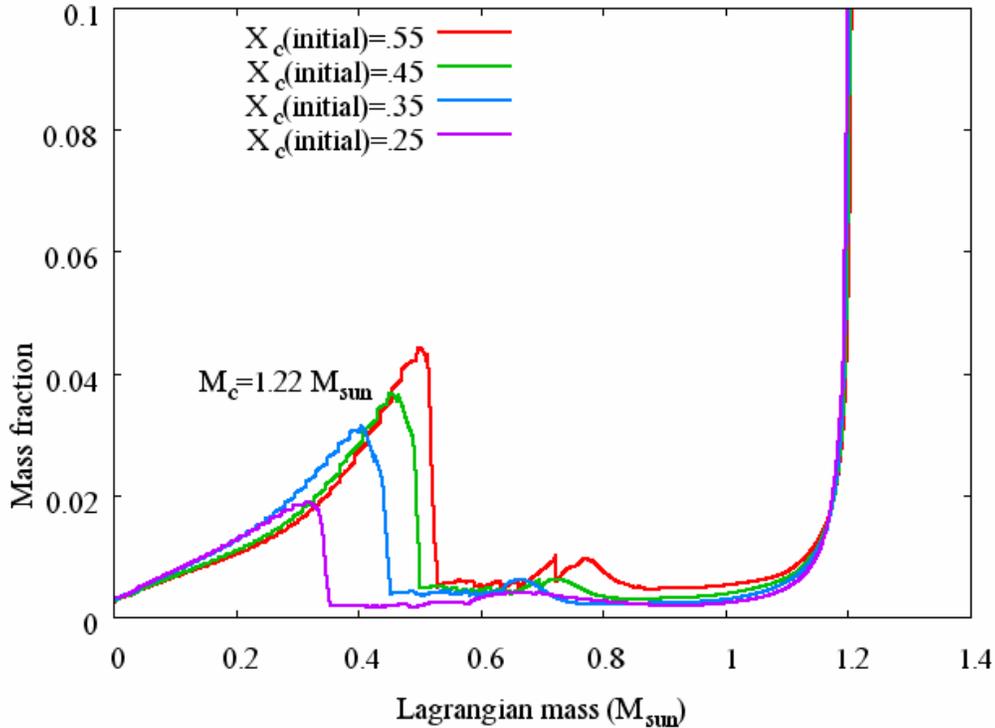

**Figure 13: The final carbon mass fraction profile in carbon star models of various initial carbon mass fractions with total mass of 1.22 M$_\odot$ after off-center carbon burning.**

The widely accepted scenario for the onset of a type I supernova explosion is accretion of matter by a white dwarf, therefore in the following we will discuss the various possibilities of accretion onto the white dwarves resulting from the evolutionary scenarios described above.

### 3.3. Accretion

For the purpose of understanding the various possibilities, we used the standard technique to investigate the outcome of accretion as a function of the relevant parameters. In a realistic case the accreted matter could be hydrogen, helium, carbon or a mixture of these. In the case of hydrogen or helium accretion, these also ignite, and form a burning shell advancing outward, leaving behind a newly-made layer of carbon – oxygen. Thus the composition of the accreted matter influences the effective rate of carbon – oxygen accretion. Variations in the entropy of the accreted matter can also be represented by dictating a suitable effective accretion rate. As we will show, the carbon mass fraction in the accreted matter can be of importance, but even then the main parameter is $\dot{M}$.

It is known that (*Nomoto & Sugimoto 1977*, *Nomoto et al. 1984*) the relevant $\dot{M}$ for a supernova has to be:

*(3-1)* $$\dot{M} \geq 7.5 \times 10^{-7} M_{\Theta} / yr$$

This rate results from the mass-luminosity relation given originally by *Paczynski 1970* for a double shell of hydrogen and helium burning:

*(3-2)* $$L/_{L_{\Theta}} = 59250 \left( M/_{M_{\Theta}} - 0.52 \right)$$

Translating this relation to $\dot{M}$ gives:

*(3-3)* $$L = q = \dot{M} X Q_{\nu}$$

Here $X$ is the hydrogen mass fraction, and $Q_{\nu}$ is the *Q-value* of hydrogen burning.

It is worth to note, that in the case of helium burning, $\dot{M}$ is expected to be, and indeed is, higher by at least* a factor of about 10, since $Q_{\nu}$ is lower.

We decided to map the results for a wide range of $\dot{M}$ :

*(3-4)* $$7.5 \times 10^{-7} M_{\Theta} / yr \leq \dot{M} \leq 3 \times 10^{-4} M_{\Theta} / yr$$

_______________________

* The luminosity itself is higher.

It is important to note, that for rates in this range, the initial mass of the white dwarf is insignificant, due to the "convergence feature" of the evolutionary paths. This argument is not to be confused with the fact, that when dealing with much smaller accretion rates, the initial mass is of much more importance.

In order to compare our results to the literature, we chose from among the many publications on this subject the extensive survey by *Nomoto & Sugimoto 1977*, followed by a study of the dependence on the accretion rate by *Sugimoto & Nomoto 1980*. It was important for us to verify that the evolutionary paths we get are similar to the literature, and thanks to the uniqueness of this path (for given accretion and neutrino loss rates) as mentioned above, this is indeed the case. It is obvious, that the onset of carbon burning depends also on the mass fraction, and we are especially interested in the cases where this mass fraction is particularly low.

Figure 14 shows the results for a model with mass of *1.18 $M_\odot$*, for various accretion rates in the range explained above.

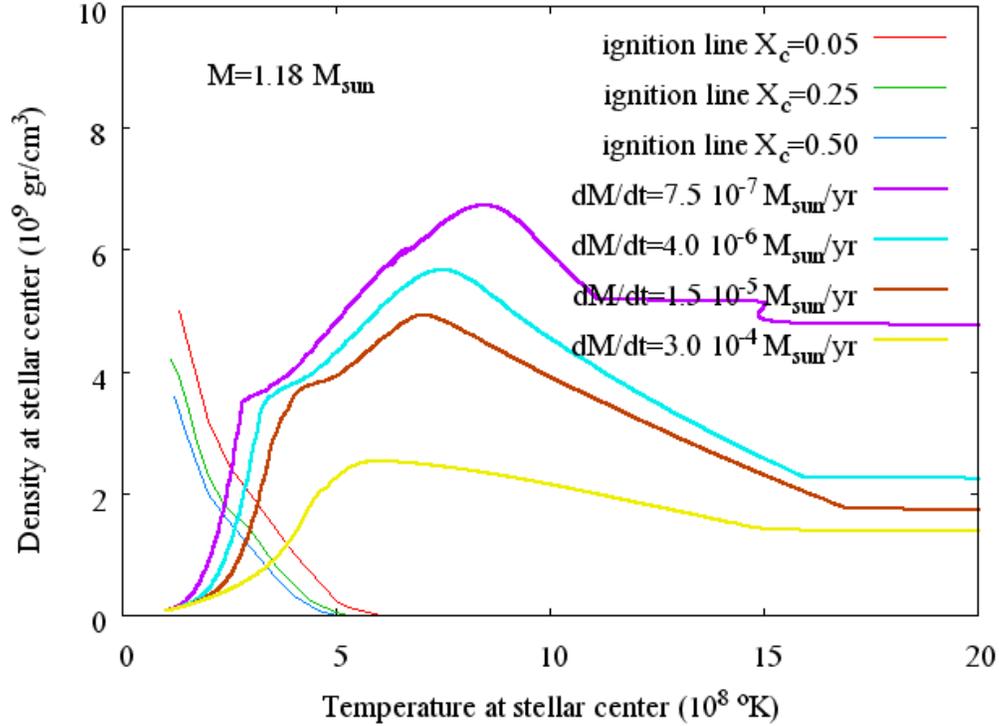

**Figure 14: The effect of accretion rate on the evolution of accreting carbon star models towards explosive ignition of carbon. All stars have an initial mass 1.18 M$_\odot$. All models have electron capture onto Mg$^{24}$ taken into account.**

As mentioned, all the above results are for accretion of matter devoid of carbon. This choice has the following reason. It turns out that, as previously mentioned, in the case of carbon accretion at high $\dot{M}$ above a certain threshold, which depends on the mass fraction of carbon in the accreted matter, carbon ignition occurs in the accreted layer, and a burning front develops, advances inward, and as already mentioned (section 3.2.2), might reach the center before explosion occurs. We didn't follow the front in this case, but we mapped the combinations of accretion rates and carbon mass fractions which lead to such a case. Figure 15 demonstrates this for a carbon – oxygen model of mass *1.12 M$_\odot$*. We can see, that for an accretion rate of *7.5 X 10$^{-7}$ M$_\odot$/yr*, carbon does not ignite in the accreted layer, even if the carbon mass fraction in the accreted matter is as high as *0.5*. For a higher accretion rate of *7.5 X 10$^{-6}$ M$_\odot$/yr*, such ignition does not occur for a carbon mass fraction of *0.01*, but does occur for *0.05* and above.

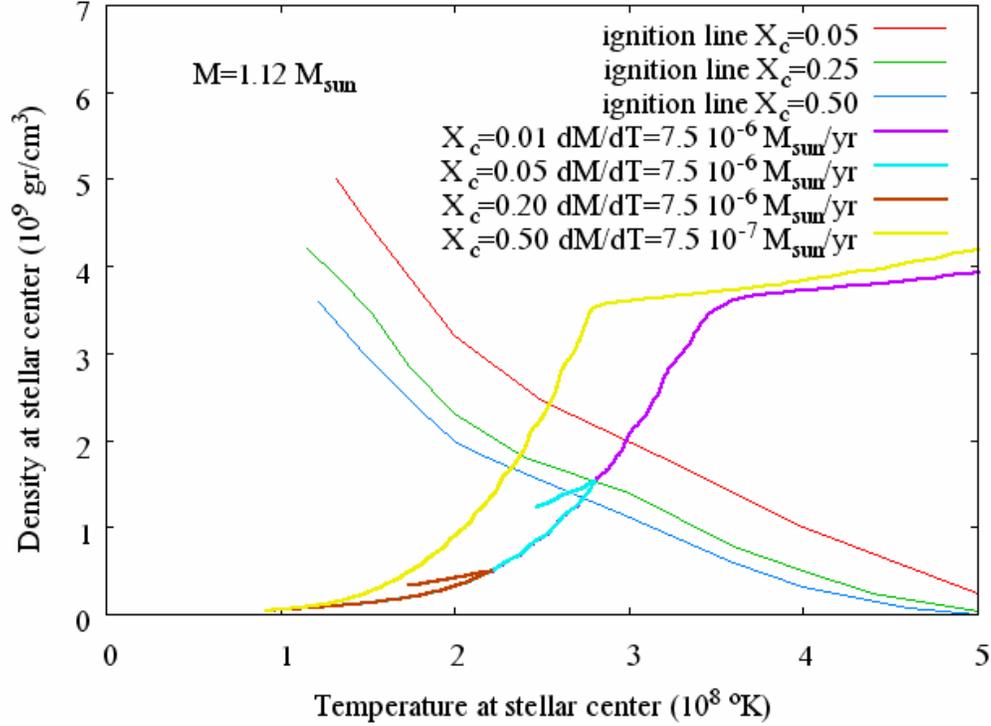

**Figure 15: The effect of the carbon mass fraction in the accreted matter, for various accretion rates, on a carbon – oxygen model of mass 1.12 M_⊕.**

### 3.4. The influence of various parameters beyond ignition

Beyond ignition a convective region is formed, which goes on growing while supplying fuel to the nuclear flame, but at the same time also inducing expansion.

Clearly, convection will continue as long as the convective turn-over time-scale $t_{convec}$ is shorter than the fuel exhaustion time-scale $t_{thermo}$. Similarly, it is clear that the transition to a dynamic regime, i.e. to a situation where assuming hydrostatic equilibrium is no longer valid, will occur when the dynamic time-scale $t_{hydro}$ (which is of order of magnitude of a characteristic length divided by the speed of sound) is no longer short compared to $t_{thermo}$.

In section 2.4 we discuss the subject of timescales in detail, including the definition of these quantities and treatment of the relevant scenarios. As we already mentioned there, our treatment implies comparing $t_{convec}$ and $t_{thermo}$, and turning convection off where $t_{convec} > \alpha\, t_{thermo}$, where $\alpha$ is a "fudge factor". Figure 16 shows the sensitivity to this factor $\alpha$, and we can see that the smaller it is (i.e. convection is turned off earlier) runaway occurs earlier, i.e. at higher $\rho_c$, and lower $T_c$, since at this stage convection causes expansion and thus hinders runaway.

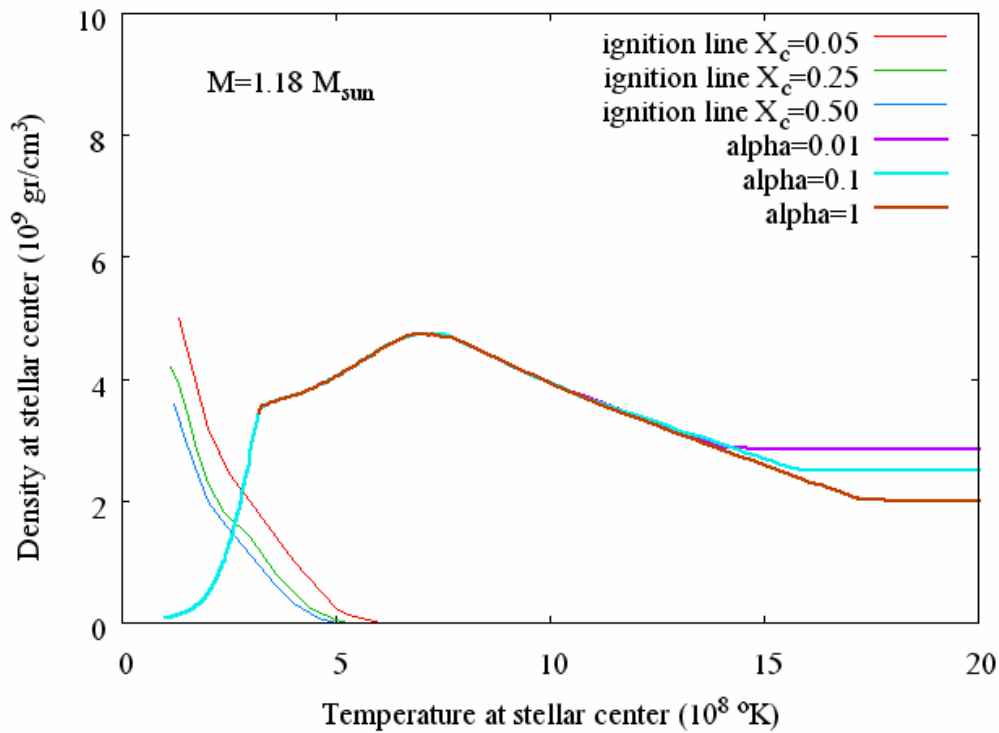

Figure 16: The influence of the criterion for stopping the convection during runaway (the parameter α as defined in section 2.4) on accreting carbon star models. All models have an initial mass 1.18 M$_\odot$ and accrete matter at a rate of 4 X 10$^{-6}$ M$_\odot$/yr.

## 3.5. The effect of electron capture processes

### 3.5.1. Electron capture on carbon burning products

Figure 17 compares the evolution with and without taking into account the effect of electron capture onto $Mg^{24}$. A considerable effect can be seen.

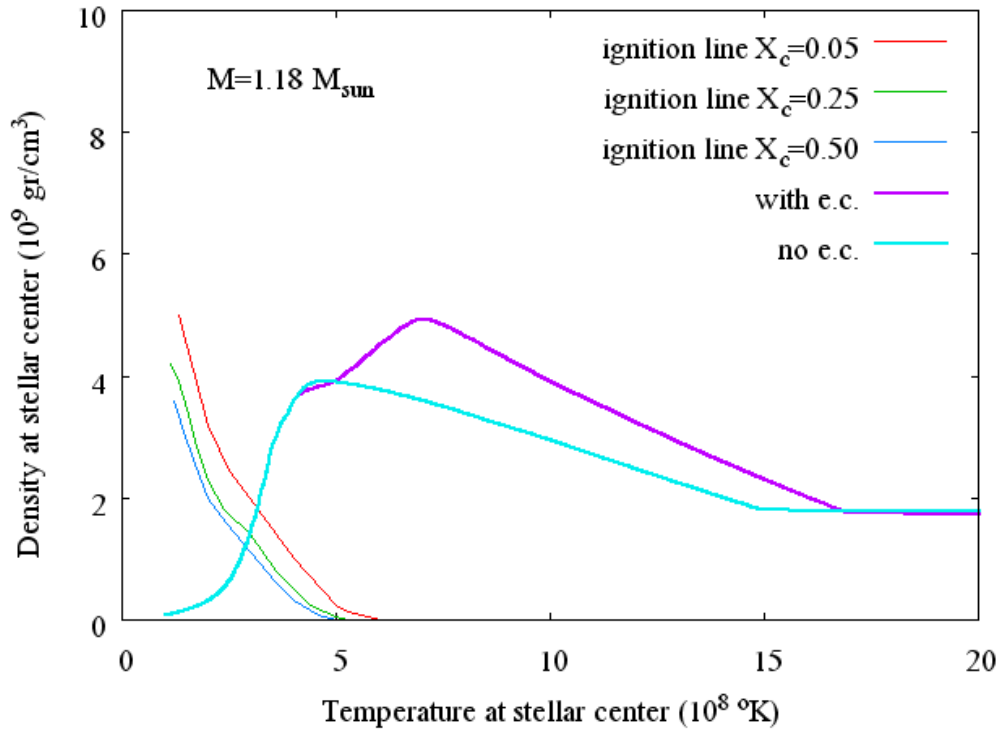

**Figure 17: The effect of electron capture onto Mg<sup>24</sup> on the evolution of accreting carbon star models towards explosive ignition of carbon. Initial mass is 1.18 M$_\odot$, the accretion rate is 1.5 X 10$^{-5}$ M$_\odot$/yr.**

Keeping in mind the existing uncertainties for this process, we checked the sensitivity of the results by modifying the mass fraction of $Mg^{24}$, which in our standard models came out to be about *0.19*. Figure 18 shows the results, and we can see that lowering the mass fraction of $Mg^{24}$ has a small effect causing earlier ignition, and thus runaway at lower $\rho_c$.

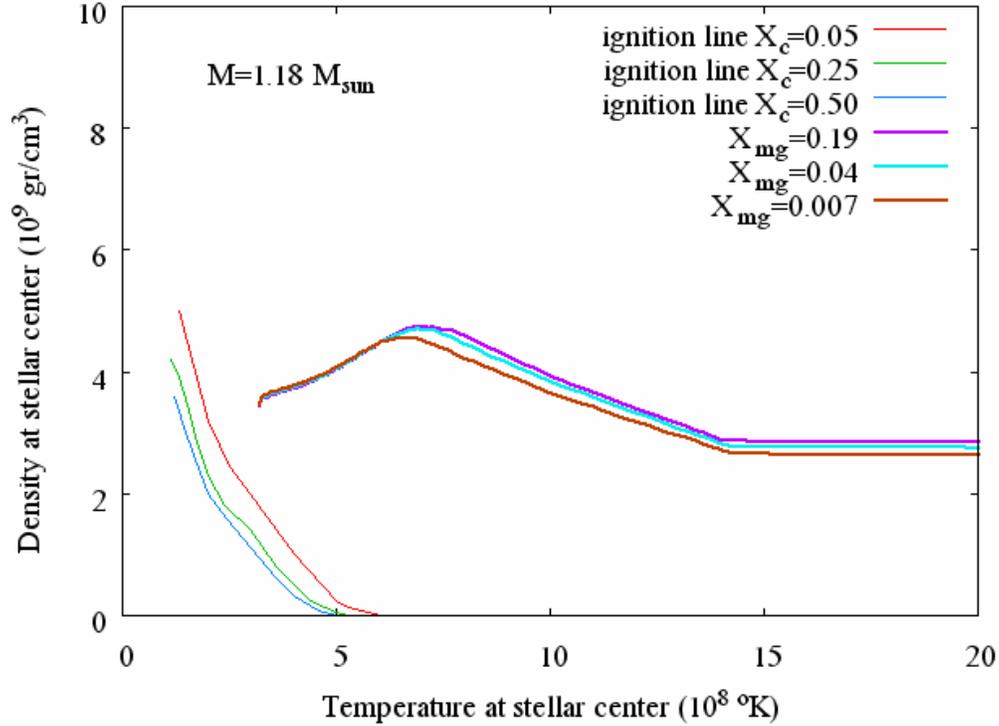

**Figure 18: The influence of the mass fraction of Mg²⁴ on the ignition and runaway of accreting carbon star models. All models have an initial mass 1.18 M⊙ and accrete matter at a rate of 4 X 10⁻⁶ M⊙/yr.**

### 3.5.2. Thermal Urca (TU)

Figure 19 shows the influence of the *TU* process for our model of *1.18 M⊙*, for various mass fraction of the *Urca* nucleus *Na²³*. We get a minor perturbation only, which is manifested as a deviation in the $\rho_c$, $T_c$ path to lower temperatures due to the local cooling. As contraction continues, the *US* moves outward together with the local heat sink. When the *US* is far enough from the center, the density and temperature at the center start rising again, and shortly the path coincides with that of the *TU*-less case. Comparison with results from the literature (*Gutierrez et al. 2005*) shows a satisfying agreement.

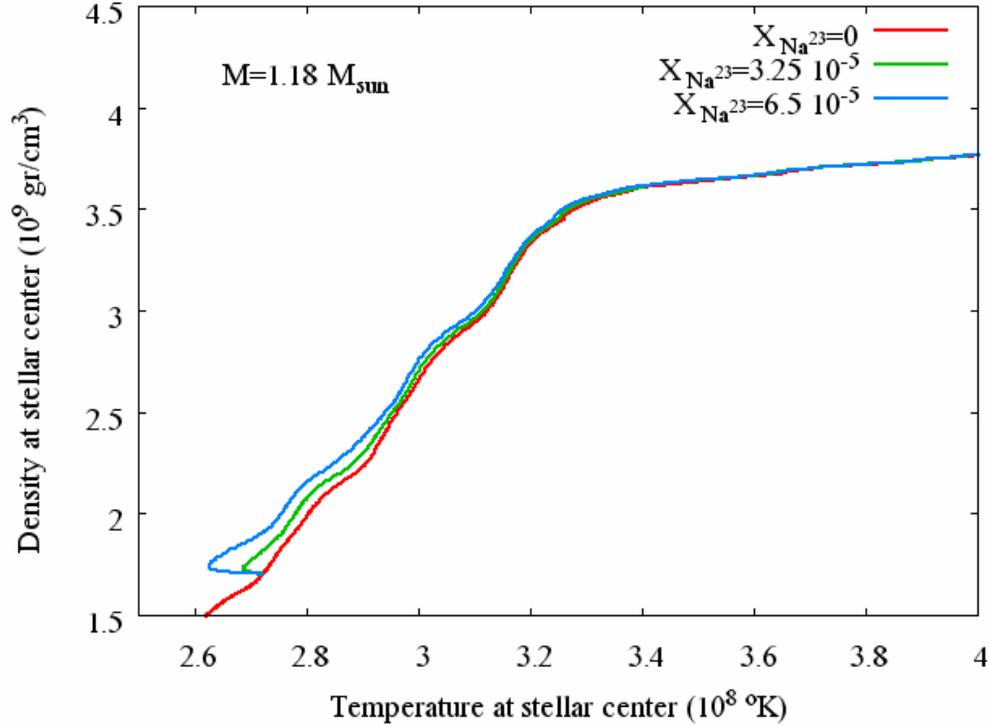

**Figure 19: The influence of the mass fraction of Na$^{23}$ for the thermal Urca process on the ignition and runaway of accreting carbon star models. All models have an initial mass 1.18 M$_\odot$ and accrete matter at a rate of 4 X 10$^{-6}$ M$_\odot$/yr.**

### 3.5.3. Convective Urca (CU)

As we already mentioned, we modeled *CU* by artificially forbidding convection above the *Urca* shell, and examined its effect by varying the Fermi energy threshold at which the *Urca* shell is located.

This works in two opposite directions. On one hand the entropy and with it the temperature increase faster (which promotes approaching *RA*), but on the other hand there is less supply of fresh fuel, which can suppress burning, and might prevent the *RA*. Clearly, lack of fresh fuel supply might have cardinal importance in a case where the mass fraction of fuel is a-priori low.

We find that a feedback mechanism exists. When the star contracts, the *US* has to move farther away from the center (due to increase in $E_f$), thus the outer boundary of the convection moves outward as well, so that the effectiveness of limiting the extent of the convective region on diminishing the fuel supply is small in all the cases we checked. It is interesting to note, that when the star goes through an expansion phase, the situation is the opposite – the *US* approaches the center, and may limit the convective region. If this limiting is significant enough while the temperature is already high enough, it might induce an earlier runaway, which will thus occur at a higher density.

Figure 20 presents a comparison between the undisturbed case, where $E_f$ peaked at about *6.26 MeV\**, with cases where convection was limited at $E_f = 4.4\ MeV$ (the threshold of $Na^{23}$), $E_f = 5.7\ MeV$ (the threshold of $Ne^{21}$), and $E_f = 5.2\ MeV$ (an intermediate value*)*.

We can see that the higher the threshold the earlier *RA* occurs, i.e. at a higher density and lower temperature. It is clear that this tendency is limited, since should we raise the threshold to the vicinity of *6.26 MeV*, the Fermi energy throughout our model will be below the threshold, and we will effectively be back to the case with no *CU* at all.

---

\* At the same time the US of $Ne^{21}$ was located at 0.09 $M_\odot$, and that of $Na^{23}$ at 0.45 $M_\odot$.

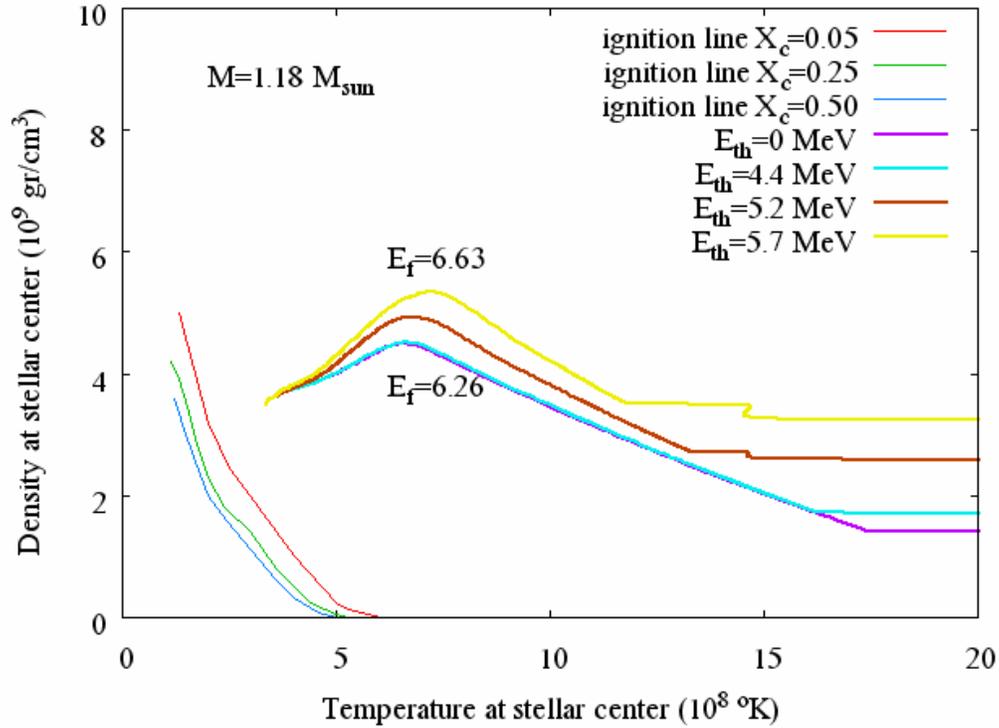

**Figure 20: The influence of the Fermi energy of the convective Urca shell on the ignition and runaway of accreting carbon star models. All models have an initial mass 1.18 M$_\odot$ and accrete matter at a rate of 4 X 10$^{-6}$ M$_\odot$/yr.**

Obviously the above conclusions are only a rough estimate, and should be in fact examined by means of multidimensional simulations, which are still difficult to undertake.

# 4. Discussion and Conclusions

Exploring stellar models which ignite carbon off-center (in the mass range of about $1.05 - 1.25\ M_\odot$, depending on the carbon mass fraction) we find that they may present an interesting *SN I* progenitor scenario, since whereas in the standard scenario runaway always takes place at the same density of about $2\ X\ 10^9\ gr/cm^3$, in our case, due to the small amount of carbon ignited, we get a whole range of densities from $1\ X\ 10^9$ up to $6\ X\ 10^9\ gr/cm^3$.

These results could contribute in resolving the emerging recognition that at least some diversity among *SNe I* exists, since runaway at various central densities is expected to yield various outcomes in terms of the velocities and composition of the ejecta, which should be modeled and compared to observations.

Several issues which were beyond the scope of this work call for further investigation:

1. A deeper treatment of the question whether thermal *Urca* can hinder the formation of a convective zone when electron capture on $Mg^{24}$ sets in is needed.

2. Our work provides initial models, which can apparently reach explosive runaway. However, our treatment of the onset of explosion, involving a very crude treatment of convection and of the convective Urca process, can only be regarded as a preliminary guideline, setting the stage for a much more profound study. The real value of our results would be judged by fitting the results of the dynamical simulations to the observational data.

3. According to *Liebert et al. 2005*, some 6% of the white dwarves have masses above $1\ M_\odot$ (and below *Chandrasekhar's* mass of $1.4\ M_\odot$). Since the carbon – oxygen stars igniting carbon off-center lie between about *1.05 – 1.18 $M_\odot$*, i.e. span about ¼ of the above range, we can roughly estimate their incidence among the white dwarf population should be in the order of magnitude of 1%. Of course, a sounder estimate of the occurrence of this type of *SNe* needs a more thorough investigation, including modeling of binary evolution.

We would like to thank the referee for his illuminating remarks.